\documentclass[showpacs,superscriptaddress,prb,preprint,11pt]{revtex4-1}%
\usepackage{graphicx}
\usepackage{amsmath}

\begin{document}
\title{First-principles investigation of magnetism and electronic structures of substitutional $3d$ transition-metal impurities in bcc Fe }
\begin{center}
\textbf{PHYSICAL REVIEW B \textbf{81}, 184423 (2010) }
\end{center}
\author{Gul Rahman and In Gee Kim\email{igkim@postech.ac.kr}}
\affiliation{Graduate Institute of Ferrous Technology, Pohang University of Science and Technology, 
Pohang 790-784, Republic of Korea}
\author{H. K. D. H. Bhadeshia}
\affiliation{Graduate Institute of Ferrous Technology, 
Pohang University of Science and Technology, 
Pohang 790-784, Republic of Korea} 
\affiliation{Department of Materials Science and Metallurgy, University of Cambridge,
Cambridge CB2 3QZ, U. K.}
\author{Arthur J. Freeman}
\affiliation{Department of Physics and Astronomy, Northwestern University, 
Evanston, IL 60208}


\begin{abstract}
The magnetic and electronic structures of $3d$ impurity atoms from Sc to Zn in 
ferromagnetic body-centered cubic iron are investigated using 
the all-electron full-potential linearized augmented plane-wave  method based on 
the generalized gradient approximation (GGA). 
We found that in general, the GGA results are closer to the experimental values than those of 
the local spin density approximation.
The calculated formation enthalpy data indicate the importance of a systematic study on 
the ternary Fe-C-$X$ systems rather than the binary Fe-$X$ systems, in steel design.
The lattice parameters are optimized and the conditions for spin polarization at the impurity sites are 
discussed in terms of the local Stoner model. 
Our calculations, which are consistent with previous work, imply that the local spin-polarizations at 
Sc, Ti, V, Cu, and Zn are induced by the host Fe atoms. 
The early transition-metal atoms couple antiferromagnetically, 
while the late transition-metal atoms couple ferromagnetically, to the host Fe atoms. 
The calculated total magnetization ($M$) of bcc Fe is reduced by impurity elements from Sc to Cr 
as a result of the antiferromagnetic interaction, with the opposite effect for solutes which couple ferromagnetically.
The changes in $M$ are attributed to nearest neighbor interactions, mostly between the impurity and host atoms. 
The atom averaged magnetic moment is shown to follow generally the well-known Slater-Pauling curve, 
but our results do not follow the linearity of the Slater-Pauling curve. 
We attribute this discrepancy to the weak ferromagnetic nature of bcc Fe. 
The calculated Fermi contact hyperfine fields follow the trend of the local magnetic moments. 
The effect of spin-orbit coupling is found not to be significant although it comes into prominence at 
locations far from the impurity sites.
\end{abstract}
\maketitle

\section{Introduction}
\label{sec:intro}
Each of the three allotropes of iron possess interesting magnetic properties 
which have a profound influence on their stability.\cite{ref-allotropy} 
The body-centered cubic (bcc) form is ferromagnetic;\cite{ref-bcc} 
the face-centered cubic (fcc) Fe at low temperatures is antiferromagnetic,\cite{ref-fcc} 
but its higher energy ferromagnetic state can be thermally excited with complex variations in magnetic structure 
as a function of temperature.\cite{fcc-noncol,fcc-noncol2} 
Ferromagnetism is eliminated when bcc iron transforms into the hexagonal close-packed (hcp) crystal structure 
at high pressures $\sim 29.5$\,{GPa}.\cite{ref-hcp}

Density-functional theory has been proven to be reliable in estimating the magnetic properties of 
iron-based transition-metal alloys, for example the pioneering work by Akai \textit{et al.}\cite{ref-akai} 
who used the Korringar-Kohn-Rostoker (KKR) Green's function method based on 
the local spin density approximation (LSDA), and Anismov \textit{et al.},\cite{ref-anim} 
who used the linear muffin-tin orbital (LMTO) Green's function method to obtain a consistent picture of 
the electronic and magnetic structures of $3d$ impurities in bcc Fe. 
However, the LSDA is known to underestimate exchange effects,\cite{ref-lda,ref-lda2} 
which are crucial for determining magnetism. 
Furthermore, lattice optimization, which may affect the electronic and magnetic structures of the impurity and host,
was not taken into account in these previous calculations. 
Lattice optimization may not change the general conclusions drawn previously, 
but can give more quantitative information. 
The relative stability of $3d$ impurities in bcc Fe is also important for practical implementations in 
the steel industry which can be precisely calculated at the optimized lattice parameters and 
for this purpose, the lattice parameters of the $3d$ impurities are also optimized. 
We focus also  on the magnetic interactions between impurity and the host element 
which can give a detailed picture of the magnetism of impurities in bcc Fe. 
The purpose of the present work was to use 
the highly precise all-electron full-potential linearized augmented plane-wave (FLAPW) method\cite{FLAPW} 
based on the generalized gradient approximation (GGA)\cite{PBE} 
to study the magnetic and electronic structures and hyperfine fields of Fe 
where it is substitutionally alloyed with the $3d$ transition metals. 
The longer term aim is to be able to contribute to the development of the so-called electrical steels 
which are used in the manufacture of motors and transformers.

\section{Calculation Method}
\label{sec:method}

\begin{figure}
\includegraphics[width=0.5\columnwidth]{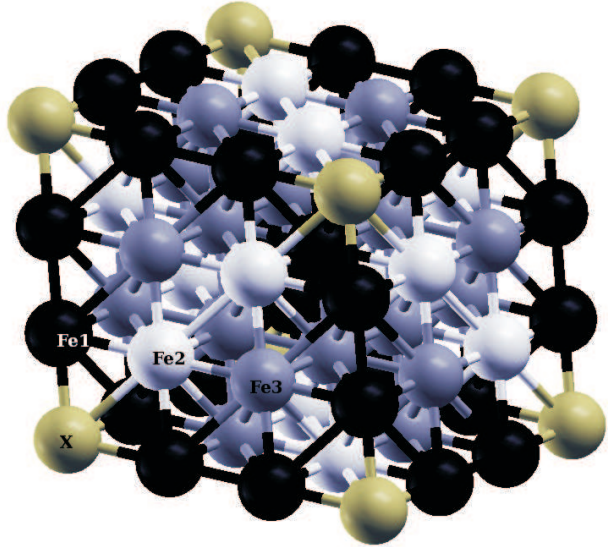}
\caption{(Color online) The model of $3\times 3\times3$ supercell of the primitive cell of bcc Fe, which contains 27 atoms. 
The Fe1, Fe2, and Fe3 atoms are represented by black, white, and grey spheres, respectively, 
while the impurity atom (corner atom) is marked with X.} \label{supercell}
\end{figure}

A $3\times 3\times 3$ supercell of the primitive bcc cell containing 27 Fe atoms (Fig.~\ref{supercell}) 
was used with $3d$ solute atoms ($X= \mathrm{Sc} - \mathrm{Zn}$) substituting a body-centered iron atom, 
giving a composition $X_1$Fe$_{26}$ equivalent to a concentration of $3.7$\;{at.\%}, 
which is consistent with the levels of solute added to steels. 
The impurity atom $X$ has first, second and third nearest neighbors designated Fe1, Fe2, and Fe3 respectively.

The Kohn-Sham equations\cite{DFT} were solved in the framework of the FLAPW\cite{FLAPW} method under 
the GGA.\cite{PBE} 
An energy cutoff at $4$\,{$(2\pi/a)$}, where $a$ is the lattice parameter of each calculation, 
employed for expanding the Linearized Augmented Plane Wave (LAPW) basis set, 
corresponding to $\sim 2350$\,{LAPW}s per \textbf{k}-point and spin. 
A $16.1245$\,{$(2\pi/a)$} cutoff was used for the star functions depicting 
the charge density and potential in the interstitial regions. 
Lattice harmonics with $l \leq 8$ were employed to expand the charge density, potential, and wave-functions 
inside each muffin-tin (MT) sphere of radius $2.2$\,{a.u.} for all the atoms. 
Integrations inside the Brillouin zone (BZ) were performed using the improved tetrahedron method\cite{TET} 
over a $13\times 13\times 13$ mesh within the three dimensional (3D) BZ, 
corresponding to $84$\,{\textbf{k} points} inside the irreducible wedge of the 3D-BZ. 
All core electrons were at first treated fully relativistically and valence states scalar relativistically, 
\textit{i.e.}, without spin-orbit coupling (SOC).\cite{Rela} 
For spin-orbit coupling on valence states, we employed the second variation method\cite{SOC} 
with the spin diagonal parts of the density subjected to a self-consistency loop. 
During the second variation procedure, integrations inside the 3D-BZ were done in 
the full-BZ, \textit{i.e.}, 1099\;{\textbf{k} points}. 
The explicit orthogonalization (XO) scheme was employed to ensure the orthogonality between 
the core and valence states.\cite{XO} 

All atoms were fully relaxed at each lattice volume until the atomic forces on 
each atom were less than $2$\;{mRy/a.u.} 
The equilibrium lattice constants and bulk moduli $B$ were determined by fitting the total energy and 
volume to the Birch-Murnaghan equation of states.\cite{Birch} 
Using the optimized lattice constants, further calculations were carried out in 
the spin-unpolarized and spin-polarized states with and without SOC. 
Self-consistency was assumed when the difference between input and output charge densities became  less than
$1.0\times10^{-4}$\;electrons/a.u.$^{3}$ 
Note that all the computational parameters used in the present calculations satisfy 
the convergence test.\cite{Seo2009}

\section{Results and Discussions}
\label{sec:results}

\subsection{Structural Properties}
\label{subsec:structure}

\begin{figure}
\includegraphics[width=\columnwidth]{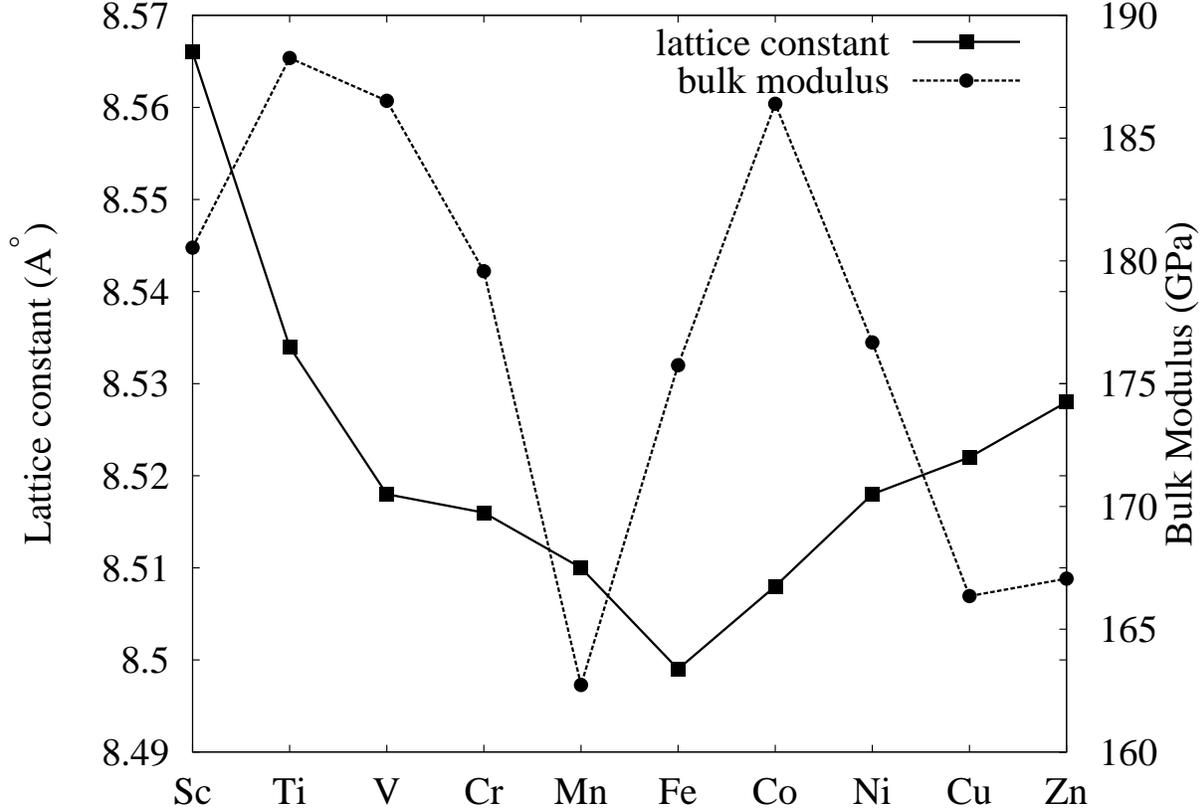}
\caption{Calculated optimized lattice constant $a$ (in units of \AA) and 
bulk modulus $B$ (in units of GPa) of $X_{1}$Fe$_{26}$. 
Filled squares (circles) on the left (right) axis show lattice constant (bulk modulus).} \label{latticestructure}
\end{figure}

Table~\ref{lattice} and Fig.~\ref{latticestructure} show the optimized lattice parameters $a$ (in units of \AA) and 
bulk moduli $B$ (in units of GPa) of $X_1$Fe$_{26}$ supercell at zero kelvin. 
The optimized lattice constant of pure bcc Fe is $2.83$\,{\AA}, 
which is only $1.4$\,{\%} smaller than 
that of the finite temperature experimental value $2.87$\,{\AA}.\cite{mm-fe} 
It is interesting to find out that the all the substitutional $3d$ impurities in bcc Fe increases the lattice parameter.  
The calculated bulk modulus of bcc Fe, $175.5$\,{GPa} is also comparable with 
the experimentally observed value.\cite{B-Modulus} 
We can see $B$ has minimum when the $d$ band is half filled,\textit{i.e.}, Mn. 
We also analyzed the local lattice expansion/contraction around the $X$ impurities and we arrive 
at a conclusion that the lattice distortions around the $X$ impurities are negligible. 
However, one can easily calculate the bond lengths of Fe-$X$, using the lattice parameters given in Table~\ref{lattice}.

\begin{table}
\caption{Calculated lattice parameters $a$ (in units of \AA), bulk moduli $B$ (in units of GPa), 
and formation enthalpy per atom ($\Delta H$ in units of {eV/atom}) of $X_1$Fe$_{26}$. 
Numbers in the parentheses are the formation enthalpy in units of {kJ/atom-mol}.}
\label{lattice}
\begin{center}
\begin{tabular}{cccccccccccccccccccccccccccccc}
\hline\hline
$X$ & Sc & Ti & V & Cr & Mn & Fe & Co & Ni & Cu & Zn\\
\hline
$a$& $\;\;\;\;8.556$ & $\;\;\;\;8.534$ &$\;\;\;\;8.518$ & $\;\;\;\;8.516$ &$\;\;\;\;8.510$ &$\;\;\;\;8.499$ &$\;\;\;\;8.508$ &$\;\;\;\;8.518$ &$\;\;\;\;8.522$ &$\;\;\;\;8.528$ &\\
$B$& $180.54\;$ & $188.26\;$ &$186.52\;$ &$179.58\;$ &$162.73\;$ &$175.75\;$ &$184.40\;$ &$176.68\;$ &$166.35\;$ &$167.06\;$ &\\
$\Delta H$& $\;\;\;\;0.034$ & $\;\;-0.011$ &$\;\;-0.004$ &$\;\;\;\;0.018$& $\;\;\;\;0.038$ &$\;\;\;\;0.000$ &$\;\;\;\;0.020$ & $\;\;\;\;0.029$ &$\;\;\;\;0.053$ &$\;\;\;\;0.038$& \\
\phantom{} & $\;\;\;\;(3.26)$ & $\;\;(-1.04)$ &$\;\;(-0.40)$ &$\;\;\;\;(1.71)$& $\;\;\;\;(3.68)$ &$\;\;\;\;(0.00)$ &$\;\;\;\;(1.91)$ & $\;\;\;\;(2.83)$ &$\;\;\;\;(5.14)$ &$\;\;\;\;(3.68)$& \\
\hline\hline
\end{tabular}
\end{center}
\end{table}

The relative stability of $X$ in bcc Fe can be understood through the formation enthalpy $\Delta H$ per atom of $X_1$Fe$_{26}$, which was calculated as follows:
\begin{equation}
\label{formation}
\Delta H=\frac{ H\left( {X}_{n}\mathrm{Fe}_{m} \right)- m H \left( \mathrm {Fe} \right)- n H \left( {X} \right)}{m+n}
\end{equation}
where $H({X}_{n}\mathrm{Fe}_{m})$ is the enthalpy of $X_n$Fe$_{m}$ for $m=26$ and $n=1$, and $H(\mathrm{Fe})$ and $H({X})$ are the total energy/atom of bcc Fe and $X$ at their ground state structures, respectively. 
For $\Delta H$, the optimized lattice parameters of the impurity $X$ crystals were also calculated, 
\textit{e.g.}, bcc Cr ($2.88$\,{\AA}) and hcp Co ($a= 2.51$\,{\AA}, $c= 4.0$\,{\AA}), 
which are close to the experimental values. 
Using the optimized lattice parameters, $\Delta H$ was calculated by the above equation and 
the results are shown in Table~\ref{lattice}.
Note that $\Delta H$ of a system is nothing more than the total energy of the system 
at zero pressure and zero Kelvin at the corresponding equilibrium lattice parameter and
is an enthalpy change for the sysnthesis of the composition from the component elements.
It is interesting to find that $\Delta H$ values of all the $3d$ elements are positive, except the Sc and V cases. 

A category of the solution atoms in Fe-C system is available:\cite{Bhadeshia2006} 
Ni and Cu are considered as the elements which enter only the ferrite phase, 
while Ti, V, Cr, and Mn are considered as the elements which form stable carbides and also enter the ferrite phase.
It seems in this Fe-$X$ system that there is no strong significant relationship between 
$\Delta H$ and the solubilities of the elements.
Many noble properties found during the alloy design for steels are able to be contributed by the characters of 
the ternary Fe-C-$X$ systems rather than the binary Fe-$X$ ones, in thermodynamic point of view.

\subsection{Local Stoner Criterion}
\label{subsec:nonmagnetic} 

\begin{figure}
\includegraphics[width=0.65\textwidth]{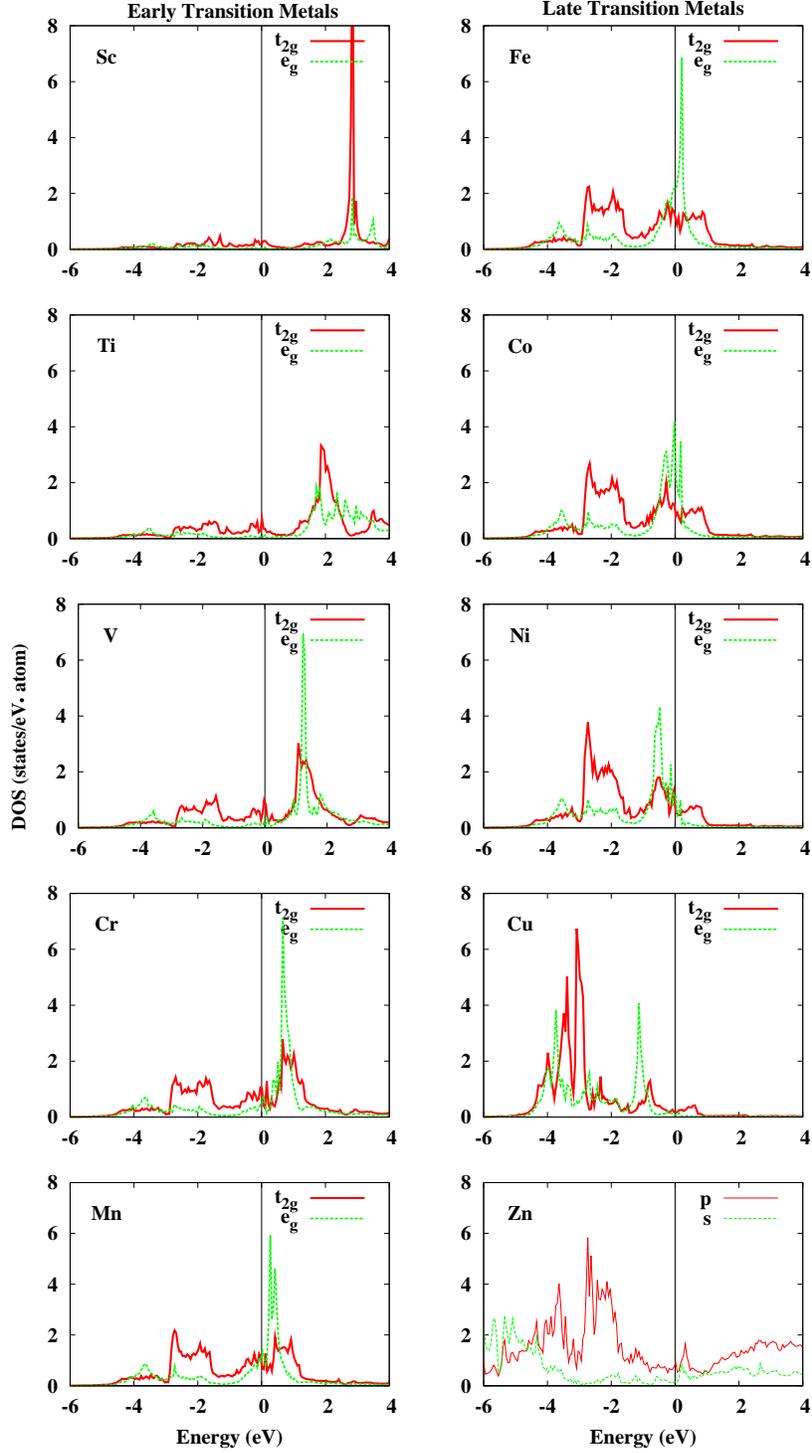}
\caption{(Color online) Calculated spin-unpolarized impurity-atom-projected local density of states of $X_{1}$Fe$_{26}$. 
Solid (dotted) lines represent the $t_{2g}$ ($e_{g}$) states, 
whereas the thin solid (dotted) lines show the $s$($p$) states, which are multiplied by a factor $20$, 
of Zn impurity at the left bottom. 
First (second) column shows early (late) transition metals. 
The Fermi energy ($E_\mathrm{F}$) is set to zero.} \label{nm-ldos}
\end{figure}

It is known that bcc Fe is a ferromagnetic metal which will be shown using spin-polarized (magnetic) calculations. 
The spin-unpolarized (non magnetic) state of bcc Fe is higher in energy than the magnetic one, 
but a knowledge of spin-unpolarized calculstions is necessary to find out the condition for 
the formation of local magnetic moment at the impurity site using the Stoner crirerion.\cite{stoner} 
The calculated spin-unpolarized impurity-site-projected local density of states (LDOS) for the solutes in 
bcc Fe are shown in Fig.~\ref{nm-ldos}, 
where the contribution from the $d$ states are decomposed into the $e_{g}$ and $t_{2g}$ states. 
The Fermi levels ($E_\mathrm{F}$) were set to zero. 
Pure bcc Fe exhibits the typical three-peak structure of $3d$ bcc metals. 
The positions of the bonding and antibonding states relative to $E_\mathrm{F}$ depend on 
the number of electrons of the impurity atoms. 
When a Sc atom substitutes for a centered Fe atom, 
the lowest lying unoccupied $d$ states are the $t_{2g}$ states. 
The addition of one valence $3d$ electron, \textit{i.e.}, when the impurity Sc is replaced by Ti, 
causes the unoccupied $t_{2g}$ states to shift towards $E_\mathrm{F}$. 
For Sc and Ti impurities the lowest lying unoccupied states are mainly the $t_{2g}$ states. 
In contrast, with V the corresponding unoccupied states become sharp and are dominated by the $e_{g}$ states.

For the other impurities, the antibonding states form virtual bound states (VBS)\cite{ref-vbs} near $E_\mathrm{F}$,
are dominated by the $e_{g}$ states. 
The $3d$ impurities also affect the LDOS of the neighboring atoms through bonding with the host atoms. 
For pure bcc Fe, one can see the dominant Fe1-$e_{g}$ states just above $E_\mathrm{F}$. 
On adding $3d$ electrons using appropriate solutes, 
these Fe1 $e_{g}$ states become narrow and their density is changed. 
After the $d$ bands are filled, 
the $p$ electrons will hybridize with the host Fe and this happens with the Zn impurity, 
where one can expect the $sp$-$d$ hybridization rather than the $d$-$d$ hybridization.

\begin{figure}
\includegraphics[width=0.8\textwidth]{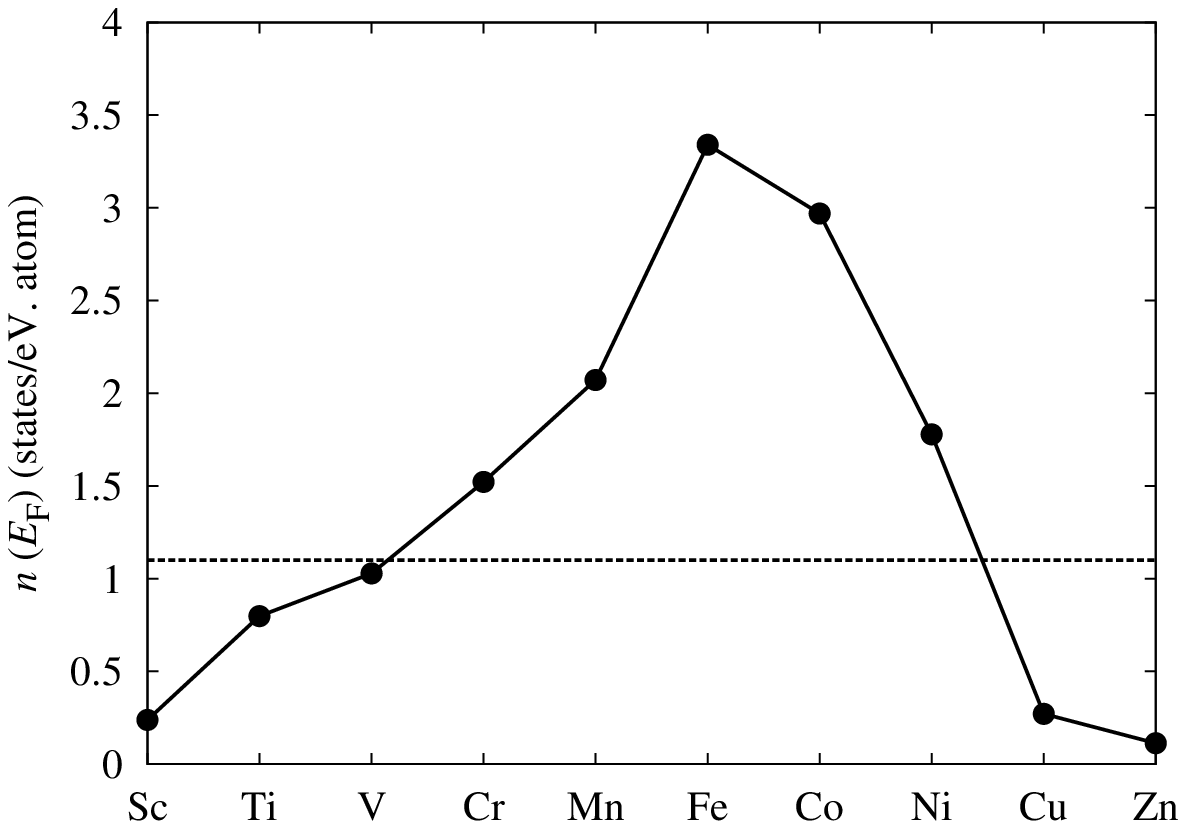}
\caption{Calculated local density of states at the Fermi energy, $n(E_\mathrm{F})$, 
for the $3d$ impurities in nonmagnetic bcc Fe. 
The dotted horizontal line represents the Stoner limits for the local spin-polarization.} \label{stoner}
\end{figure}

Figure~\ref{stoner} we show the calculated $X$ atom projected LDOS at the Fermi energy, 
denoted by $n(E_\mathrm{F})$. 
The condition for the formation of local spin polarization at the $X$ atom in bcc Fe can be 
determined approximately by adapting the Stoner criterion,\cite{stoner} \textit{i.e.}, 
$In(E_\mathrm{F}) > 1$, where $I$ is the well known atomic exchange parameter equal to 
$0.925$\,{eV} for bcc Fe.\cite{stoner-val} 
The critical value of $n(E_\mathrm{F})$ above which an intrinsic local magnetic moment arises on 
the impurity atom in bcc Fe was estimated to be $\sim 1.081$\;{states/eV$\cdot$atom}, 
shown in Fig.~\ref{stoner} as a dotted horizontal line. 
It is found that the condition for the local spin polarization is not satisfied for the Sc, Ti, V, Cu, and Zn impurities. 
In other words, the calculated local magnetic moments of impurity atoms which do not satisfy
the local Stoner criterion, are induced by the surrounding magnetic Fe atoms.

The elements from Cr to Ni, which satisfy the local Stoner criterion, 
induce the $t_{2g}$ and $e_{g}$ states to move closer to $E_\mathrm{F}$. 
For the early $3d$ impurities, $n(E_\mathrm{F})$ is mainly contributed by the $t_{2g}$ states, 
while for the late $3d$ transition-metal impurities $n(E_\mathrm{F})$ is contributed by the $e_{g}$ states. 
Once these $3d$ impurity bands cross $E_\mathrm{F}$, \textit{i.e.}, 
when the $d$ bands are completely filled, the electronic structures and the magnetism of the impurity atoms 
will be determined by the $sp$ electrons as seen, for example, in the LDOS of Zn in Fig.~\ref{nm-ldos}.

\subsection{Magnetism}
\label{subsec:local-mm}

\begin{figure}
\includegraphics[width=0.8\textwidth]{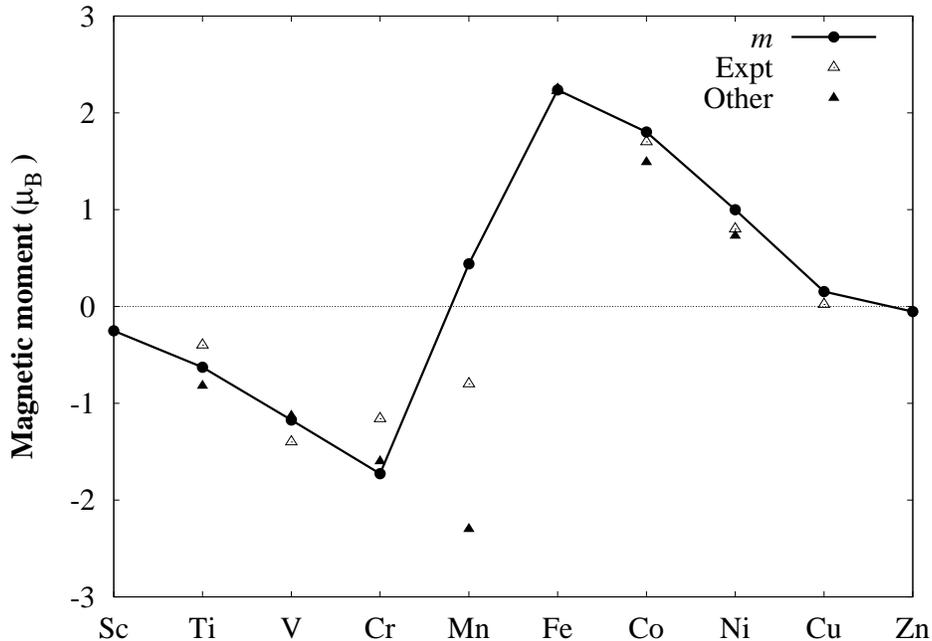}
\caption{Calculated local magnetic moment of the $3d$ impurities in bcc Fe. 
Filled circles show our calculated local magnetic moment (in units of $\mu_{\mathrm{B}}$) at the $X$ site, 
filled triangles show the previously calculated values (Other),\cite{ref-anim}  and 
open triangles show the experimental (Expt.) values taken from 
Refs.~\onlinecite{mm-ti,neutron,mm-cr,mm-fe,mm-co,mm-ni,mm-cu}.}
\label{local-MM}
\end{figure}

Figure~\ref{local-MM} shows the calculated local magnetic moment ($m$) 
within each MT sphere of the $3d$ impurity sites. 
Solutes from Sc to Cr are associated with negative values of $m$ with antiferromagnetic coupling 
with the host Fe atoms. 
It is noticeable that Sc, Ti, and V atoms have significant magnetic moments, 
even though these elements are not
magnetic elements---this is because the $3d$ magnetism, 
according to the local Stoner criterion (Fig.~\ref{stoner}), 
is not intrinsic but is induced via the host iron atoms. 
It is considered that the largely induced $m$ of the impurities from Sc to V are due to the fact that
their $3d$ states are induced.

Cr and Mn, which satisfy the Stoner criterion and are antiferromagnetically and ferromagnetically coupled to 
the iron respectively, have intrinsically large magnetic moments which are not attributed to 
the neighboring iron atoms.  
In the case of Co and Ni, the local magnetic moments are positive and are coupled ferromagnetically to the host Fe. 
In contrast to the induced $m$ of the early transition-metal impurities from Sc to V, 
the calculated $m$ for Cu and Zn impurities, which have complete $d$ shell occupation, are very small. 
Especially, the Zn impurity shows local diamagnetism. We may say that Sc to Cr might not be useful due to
antiferromagnetic coupling when designing iron alloys where a high saturation magnetization is required.

It seems that as a general principle, similar to the Hund's rules, 
impurity atoms with $3d$ states less than half-filled will tend to couple antiferromagnetically with iron and
ferromagnetic coupling occurs when the atoms have more than half-filled states, 
and this statement was also confirmed by spin-density contours. 
Figure~\ref{local-MM} also shows published data\cite{ref-anim,mm-ti,neutron,mm-cr,mm-fe,mm-co,mm-ni,mm-cu}
which seem to be consistent with the trends illustrated in Fig.~\ref{stoner}.

The case of Mn impurity attracts attention, because the local magnetic moments of Mn are very sensitive to volume.
Our analysis showed that Mn can couple antiferromagnetically as well as 
ferromagnetically depending on the lattice volume. 
For example, at the experimental lattice constant of bcc Fe ($5.4169$\,{a.u.}), 
Mn couples antiferromagnetically with the host Fe atoms, 
but couples ferromagnetically at the optimized lattice constant of Fe$_{26}$Mn ($5.3608$\,{a.u.}). 
The unstable behavior of local magnetic moments of Mn may cause disagreement 
between experiments and theoretical calculations. 
There is a significant discrepancy with the experimental data for Mn; 
spin-unpolarized neutron diffraction measurements of FeMn alloys gave values of a Mn local moment from 
$0.0 \pm 0.2$\,{$\mu_\mathrm{B}$}\cite{mm-exp2} to $1.0 \pm 0.2$\,{$\mu_\mathrm{B}$},\cite{mm-ti} 
while polarized neutron diffraction measurements give  
$0.77$\,{$\mu_{B}$} for parallel\cite{mm-exp4} and $-0.82$\,{$\mu_{\mathrm{B}}$} for
antiparallel\cite{neutron} to the orientation of $m$ of Mn. 
The previously calculated $m$ of Mn in Fe are:
$0.6$\,{$\mu_{\mathrm{B}}$},\cite{ref-akai} $1.6$\,{$\mu_{\mathrm{B}}$},\cite{mm-theory1} and  
$-2.30$\,{$\mu_{\mathrm{B}}$} and $1.6$\,{$\mu_{\mathrm{B}}$}.\cite{mm-theory2} 
The discrepancies between the individual calculations are due to the use of different computational methods and 
the exchange-correlation potentials. 
The KKR-Green's function method by Akai \textit{et al.}\cite{ref-akai} showed that the variation of $m$ with 
the impurity nonintegral nuclear charges $Z$, leads $m$ with negative values for $Z \leq 25$, 
but that the sign is reversed beyond $Z=25.17$. 
This implies that the $m$ of Mn is not only susceptible to nuclear charges but also to volumes. 
On the other hand, recent calculations indicate the noncollinear magnetic structures for FeMn.\cite {ref-kohji} 
The discrepancies between the calculated and the experimental values of $m$ of Mn shows that 
the exact experimental determination of the value of the Mn impurity magnetic moment might be complicated by 
its pronounced concentration\cite{neutron} and temperature dependence\cite{mm-ti} in FeMn alloys.


\begin{figure}
\includegraphics[width=0.8\textwidth]{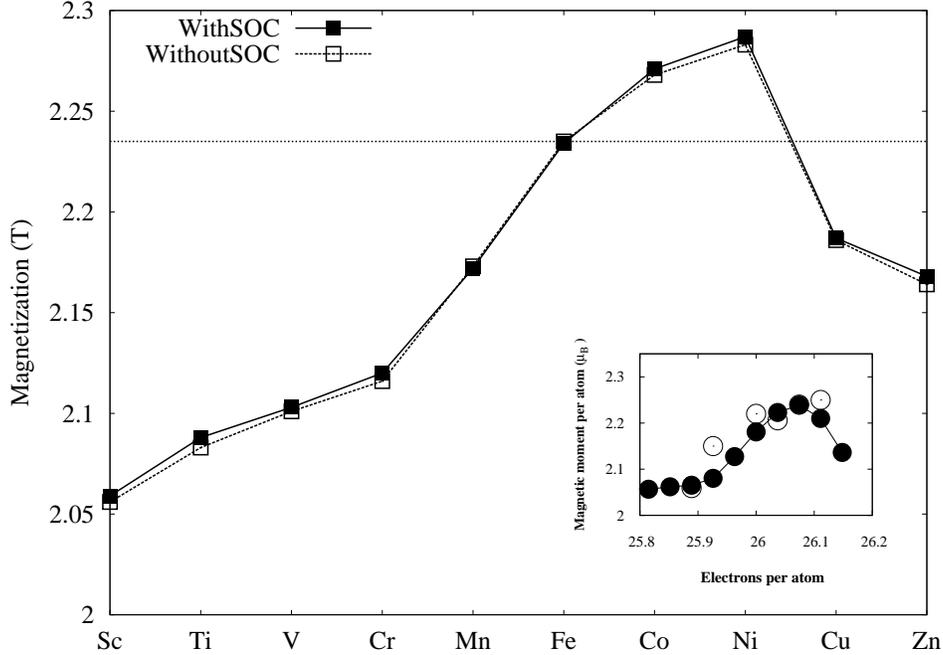}
\caption{Calculated total magnetization $M$ of $X_{1}$Fe$_{26}$ (in units of T). 
Open squares show $M$ without spin-orbit coupling (SOC) and filled squares show $M$ with SOC. 
The horizontal dotted line shows $M$ of pure Fe without SOC. 
The inset shows the atom averaged magnetic moment ($\bar{m}$) (open circles) 
in units of $\mu_{\mathrm{B}}$ without SOC versus the number of electrons per atom of $X_{1}$Fe$_{26}$. 
The filled circles show the experimental values taken from Ref.~\onlinecite{slater-val}.}
\label{Magnetization}
\end{figure}

Judging from the dependence of local magnetic moments, 
one can  expect that solutes from Sc to Cr will decrease the magnetization $M$  of bcc Fe due to 
antiferromagnetic coupling, 
while  Co to Ni will have the opposite effect based on ferromagnetic coupling to the host Fe atoms. 
This is indeed observed as shown in Fig.~\ref{Magnetization}, 
where the calculations are presented with and without SOC. 
$M$ is seen to vary linearly with the atomic number from Sc to Mn with a large increase from Mn to Fe. 
The Mn case is quite different because the sign of $m$ of Mn is very sensitive to external perturbation 
(volume in this case) around the equilibrium volume. 
Therefore, there are two competing factors that determine the change in $M$ with respect to pure bcc Fe,
\textit{i.e.}, nearest neighbor (NN) interaction and lattice volume. 
The lattice volume of $X$ (Sc--Mn) is  larger than bcc Fe, so one can expect small $M$, 
and at the same time the nearest-neighbor interactions favor the AFM coupling between 
$X$ (Sc--Cr) and Fe atoms and decreases $M$.

For Co and Ni impurities, although the lattice volume is larger but the NN interactions favor FM coupling and 
increases $M$.
From Fe to Ni, $M$ increases linearly and we already discussed the enhanced $M$ of Ni in bcc Fe due to 
the spin-flip of Fe1 $d$ states.\cite{ref-JKMAG} 
On the other hand, $M$ decreased sharply for Cu and Zn, because as stated previously, 
their $3d$ bands are fully occupied and hence are not susceptible to magnetic interactions with the host atoms. 
The calculated magnetic moment of bcc Fe is found to be $2.22$\,{$\mu_{\mathrm{B}}$}, 
the same as that observed experimentally.\cite{mm-fe} 
The calculated trend that solutes from Sc to Mn decrease the total magnetization of the system 
has been observed experimentally.\cite{ref-mag,ref-v,ref-femn}  
Co and Ni enhance the magnetization of Fe and this compares well with 
experimental measurements.\cite{mm-co,mm-ni,mm-co2}

Before going into the microscopic description of the magnetic interaction of an $X$ impurity with 
the neighboring Fe atoms, our calculated trends of $M$ can be compared with 
the Slater-Pauling curve,\cite{slater,pauling} 
which is a plot of atom averaged magnetic moment ($\bar{m}$) of ferromagnets versus the electron-to-atom ratio. 
Here, we used $\bar{m}$ to distinguish it from the magnetization $M$. 
Bcc iron as a weak ferromagnet and alloys based on bcc iron can form the left branch of the Slater-Pauling curve. 
Our calculated $\bar{m}$ of $X_{1}$Fe$_{26}$ can follow the simple phenomenological relation discussed in Ref.~\onlinecite{stoner}, \textit{i.e.},
\begin{equation}
\label{left-branch} \bar{m}=m_{A}^0+x(Z_{B}-Z_{A}),
\end{equation}
for the left branch, \textit{i.e.}, $\bar{m}$ increasing branch,
\begin{equation}
\label{right-branch} \bar{m}=m_{A}^0-x(Z_{B}-Z_{A}),
\end{equation}
for the right branch, \textit{i.e.}, $\bar{m}$ decreasing branch, 
where $m_{A}^0$ is the magnetic moment of the host atom, $x$ is the solute concentration, 
and $Z_{A}$ and $Z_{B}$ are the valences of the host and the solute atoms, respectively. 
The inset of Fig.~\ref{Magnetization} shows $\bar{m}$ versus the number of electrons per atom. 
The $\bar{m}$ curve shows a local maximum at about $26.074$\,{electrons per atom}, 
which is approximately similar to the other bcc alloys on the Slater-Pauling curve. 
However, it is noticeable that the left branch does not follow the simple linear relation in Eq. (\ref{left-branch}). 
We attribute that this offset from the linear behavior is caused by the weak ferromagnetic nature of bcc Fe, 
discussed in the following.

\begin{figure}
\includegraphics[width=0.8\textwidth]{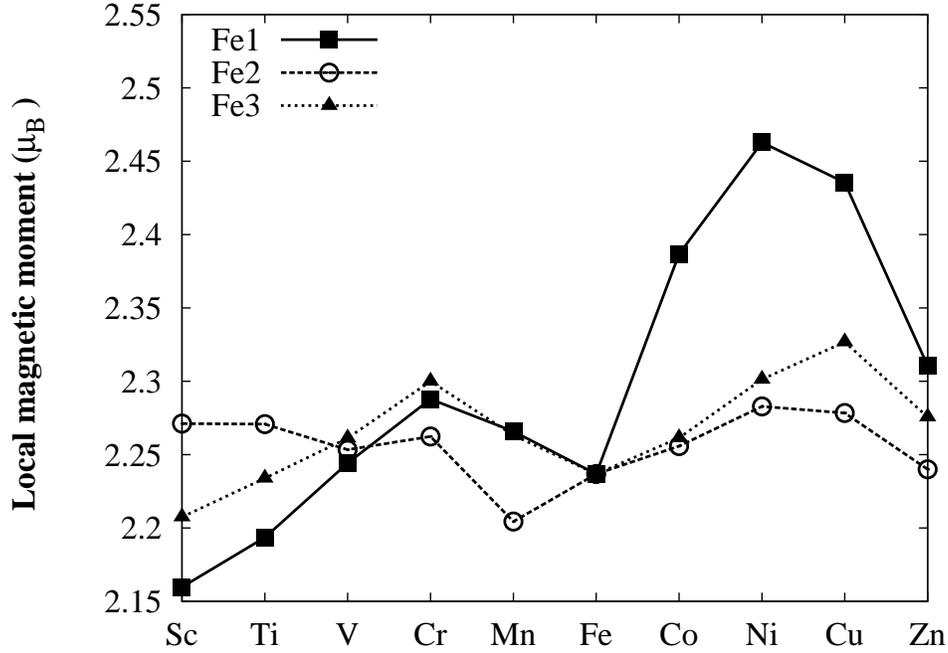}
\caption{Calculated local magnetic moments (in units of $\mu_\mathrm{B}$) of Fe1, Fe2, and Fe3 atoms. 
Filled squares and triangles represent the local magnetic moments of Fe1 and Fe3 atoms, respectively, 
whereas the open circles represent Fe2 atoms.}
\label{NN-MM}
\end{figure}

\squeezetable
\begin{table}
\caption{Calculated site-projected local spin magnetic moments 
(in units of $\mu_\mathrm{B}$) of $X_1$Fe$_{26}$ inside each muffin-tin (MT) sphere 
with and without spin-orbit coupling (SOC). 
$M$ is the total magnetization in units of $\mathrm{T}$. 
The lower portion of the table shows the effect of SOC.} \label{table1}
\begin{center}
\begin{tabular}{cccccccccccccccccccccccccccccc}
\hline \hline \multicolumn{11}{c}{Without SOC}\\
\hline
Site&Sc & Ti & V & Cr & Mn & Fe & Co & Ni & Cu & Zn\\
\hline $M$& $\;\;\;2.059$ & $\;\;\;2.088$ & $\;\;\;2.103$ &
$\;\;\;2.120$ & $\;\;\;2.172$ & $\;\;\;2.235$ & $\;\;\;2.271$ &
$\;\;\;2.287$ & $\;\;\;2.188$ &
$\;\;\; 2.168$\\
\hline
$X$& $-0.251$ & $-0.628$ & $-1.173$ & $-1.727$ & $\;\;\;0.440$ & $\;\;\;2.237$ & $\;\;\;1.802$ & $\;\;\;0.999$ &$\;\;\;0.150$ & $-0.053$\\
Fe1& $\;\;\;2.160$ & $\;\;\;2.194$ & $\;\;\;2.244$ & $\;\;\;2.288$ & $\;\;\;2.266$ & $\;\;\;2.237$ & $\;\;\;2.387$ & $\;\;\;2.463$ & $\;\;\;2.360$ & $\;\;\;2.311$\\
Fe2& $\;\;\;2.271$ & $\;\;\;2.271$ & $\;\;\;2.253$ & $\;\;\;2.262$ & $\;\;\;2.204$ & $\;\;\;2.237$ & $\;\;\;2.256$ & $\;\;\;2.283$ & $\;\;\;2.223$ & $\;\;\;2.240$\\
Fe3& $\;\;\;2.207$ & $\;\;\;2.234$ & $\;\;\;2.261$ & $\;\;\;2.300$ & $\;\;\;2.263$ & $\;\;\;2.237$ & $\;\;\;2.261$ & $\;\;\;2.301$ & $\;\;\;2.266$ & $\;\;\;2.276$\\
\hline\hline
\multicolumn{11}{c}{With SOC} \\
\hline
Site&Sc & Ti & V & Cr & Mn & Fe & Co & Ni & Cu & Zn\\
\hline $M$& $\;\;\;2.056$ & $\;\;\;2.083$ & $\;\;\;2.101$ &
$\;\;\;2.116$ & $\;\;\;2.173$ & $\;\;\;2.234$ & $\;\;\;2.268$ &
$\;\;\;2.283$ & $\;\;\;2.186$ & $\;\;\;2.164$\\
\hline
$X$ & $-0.251$ & $-0.628$ & $-1.171$ & $-1.723$ & $\;\;\;0.470$ & $\;\;\;2.235$ & $\;\;\;1.797$ & $\;\;\;0.996$ & $\;\;\;0.151$ & $-0.052$\\
Fe1 & $\;\;\;2.156$ & $\;\;\;2.189$ & $\;\;\;2.242$ & $\;\;\;2.285$ & $\;\;\;2.264$ & $\;\;\;2.235$ & $\;\;\;2.382$ & $\;\;\;2.458$ & $\;\;\;2.357$ & $\;\;\;2.306$\\
Fe2 & $\;\;\;2.267$ & $\;\;\;2.265$ & $\;\;\;2.251$ & $\;\;\;2.259$ & $\;\;\;2.205$ & $\;\;\;2.235$ & $\;\;\;2.252$ & $\;\;\;2.280$ & $\;\;\;2.221$ & $\;\;\;2.236$\\
Fe3 & $\;\;\;2.203$ & $\;\;\;2.229$ & $\;\;\;2.258$ & $\;\;\;2.295$ & $\;\;\;2.262$ & $\;\;\;2.235$ & $\;\;\;2.259$ & $\;\;\;2.298$ & $\;\;\;2.263$ & $\;\;\;2.272$\\
\hline\hline
\end{tabular}
\end{center}
\end{table}

Figure~\ref{NN-MM} and the results in Table~\ref{table1} show that it is the interaction of 
$X =$\,\,Sc--Mn with first neighbor Fe1 atoms that is most responsible for the reduction in $M$ of the alloy. 
With Cr and Mn which exhibit local intrinsic spin polarizations, the reduction in $M$ is caused mainly by 
the interaction with the Fe2 atoms. The importance of Fe2 can be understood in terms of 
the local symmetry of the Cr and Mn atoms in bcc Fe. 
The width of the splitting of the bonding and antibonding states usually depends on the spatial separation of 
the atoms, and becomes large if the atoms are close together. 
The Cr and Mn atoms, surrounded by the nearest eight Fe1 atoms, 
are located at the corners of a cube at a distance of $\sqrt{3}a/2$. 
The atomic wave-functions of Cr or Mn-$e_{g}$ and Fe1-$t_{2g}$ overlap strongly and form hybrid orbitals. 
The impurity atoms are also surrounded by six next nearest neighboring Fe2 atoms, 
which form an octahedral cage around Cr and Mn atoms at a distance of $a$. 
Consequently, the hybridization is smaller and hence so is the resulting splitting of the hybrid orbitals. 
Due to the weak antiferromagnetic interaction between the Cr (or Mn) and the Fe2 atoms, 
the magnetic moment of the Fe2 atom is smaller than that of the Fe1 atom.

The variation in $m$ of Fe1, Fe2, and Fe3 atoms is not monotonic for solutes with $d > 5$. 
The Fe1 atoms have a much larger $m$ than those of Fe2 and Fe3 atoms for the late transition-metal solutes. 
The enhancement of magnetization due to the late transition-metal impurities in bcc Fe is caused mainly by 
the Fe1 atoms. 
The origin of such an enhanced magnetization in bcc Fe was already discussed in Ref.~\onlinecite{ref-JKMAG} 
where we found strong interactions between the Fe1-$t_{2g}$ and Ni-$e_{g}$ states, through the VBS.


\begin{figure}
\includegraphics[width=0.65\textwidth]{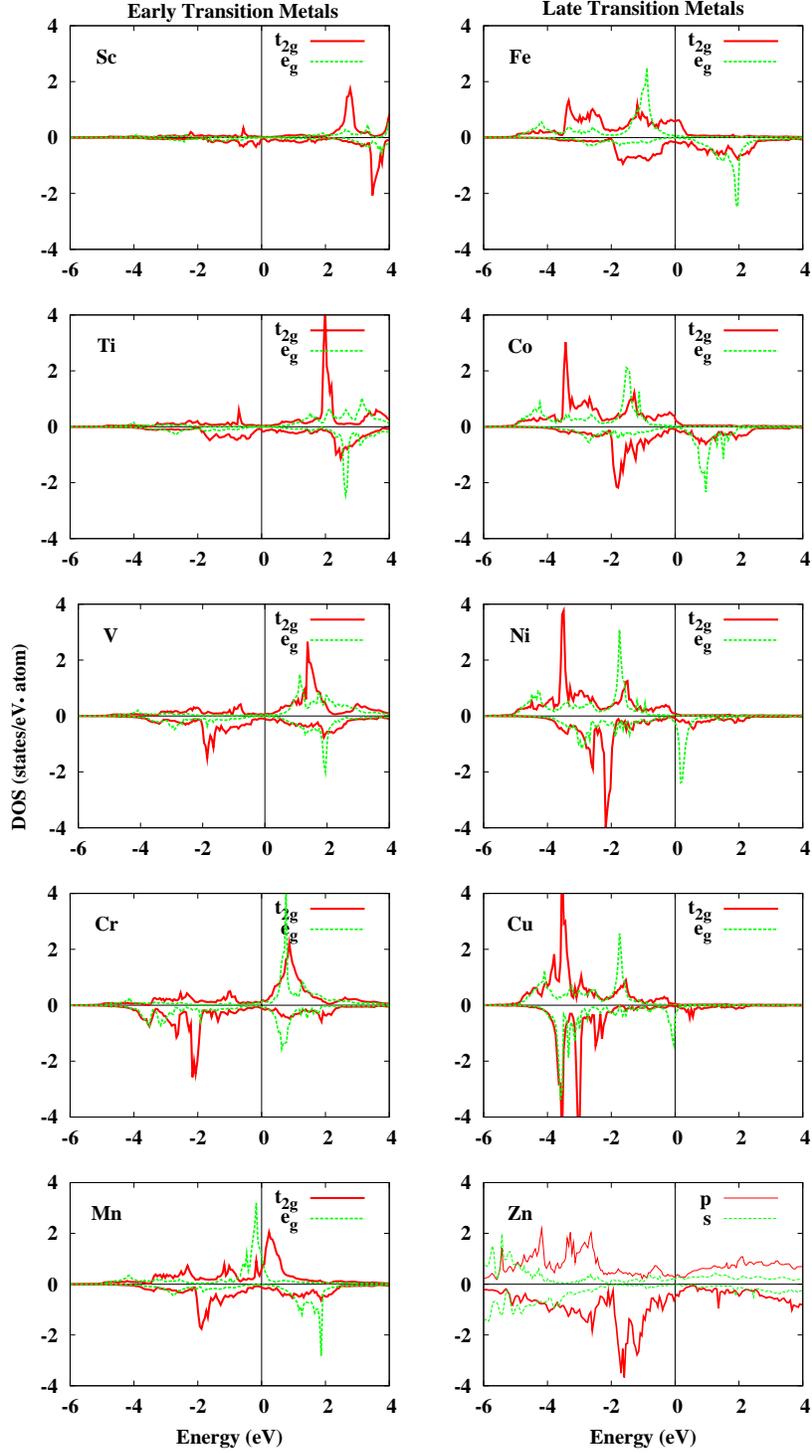}
\caption{(Color online) Calculated spin-polarized impurity-site-projected local density of states of $X_{1}$Fe$_{26}$. 
The upper (lower) panels show majority (minority) spin states. 
Solid (dotted) lines show the $t_{2g}$ ($e_{g}$) states, 
whereas for the Zn impurity at the left  bottom the thin solid  and dotted lines show the $s$ and $p$ states, 
which are multiplied by a factor $20$. First (second) column shows early (late) transition metals. 
The Fermi levels ($E_\mathrm{F}$) are set to zero.} \label{FMspin}
\end{figure}

For further insight, it is useful to compare the spin-unpolarized LDOS in Fig.~\ref{nm-ldos} and 
the spin-polarized one in Fig.~\ref{FMspin}. 
Considering bcc Fe, or the Fe impurity case, the $e_{g}$ majority spins are almost completely occupied, 
but the $t_{2g}$ majority spins are partially filled. Due to this latter property, 
bcc Fe is so-called a weak ferromagnet.\cite{book-mohan} 
This makes the magnetic moment of bcc Fe sensitive to perturbations due to $X$ impurities.

With the substitution of Sc, the valence $3d$ electron is accommodated in the minority $t_{2g}$ states; 
if instead Ti is added, its additional $3d$ electron occupies the minority $t_{2g}$ states with both the latter and
$e_{g}$ states shifting towards lower energy. 
Thus, by adding electrons, the unoccupied states move closer to $E_\mathrm{F}$. 
Therefore, the spin-polarized LDOS indicates that for the impurities from Sc to Cr all the $3d$ electrons occupy in 
the local minority spin bands to achieve charge neutrality and consequently 
to align the impurity magnetic moments in an opposite sense to the host magnetic moments. 
For Mn impurity, the majority $e_{g}$ spin staes also become occupied. 
We verified that the local DOS at Mn site depends on the volume.

The $t_{2g}$ minority spin states are filled beyond Mn so that further additional  $3d$ electrons are forced 
to enter the $t_{2g}$ majority spin states and couple ferromagnetically to the host Fe atoms. 
It follows that from Mn to Cu the impurity magnetic moments are parallel to those of the host, 
whereas for Sc to Cr they are coupled antiparallel to the Fe host. 
The transition from antiferromagnetic to ferromagnetic coupling occurs when the VBS 
in the majority spin band crosses $E_\mathrm{F}$, \textit{i.e.}, 
the majority spin $d$ bands being occupied\cite{ref-friedel} (see Fig.~\ref{nm-ldos} and Fig.~\ref{FMspin}).

The effects of $3d$ impurities on the electronic structures of Fe1, Fe2, and Fe3 atoms (not shown here) 
are generally that the minority spin $e_{g}$ states at $\sim 2.0$\;{eV} above $E_\mathrm{F}$ are not 
strongly affected by the $3d$ impurity atoms. 
The hybridization between $X$ and the host Fe become important when the $d$ bands of impurity atoms are 
more than half filled, \textit{e.g.}, 
see the Ni and Cu where the additional peaks at Fe1 are caused by the hybridization of the $X$ impurities with 
the host Fe, through the location of the VBS.


\begin{figure}
\includegraphics[width=0.4\textwidth]{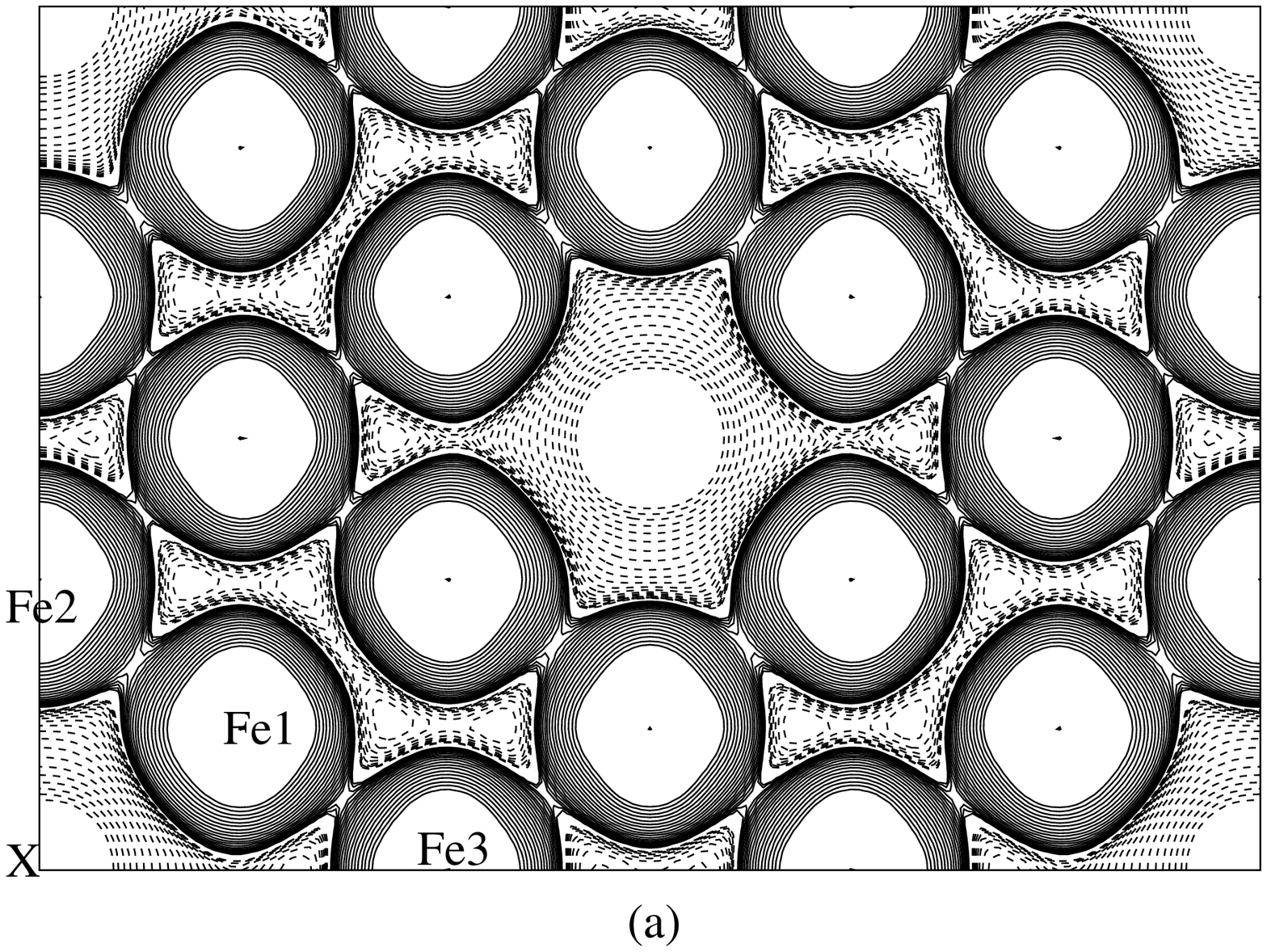}
\includegraphics[width=0.4\textwidth]{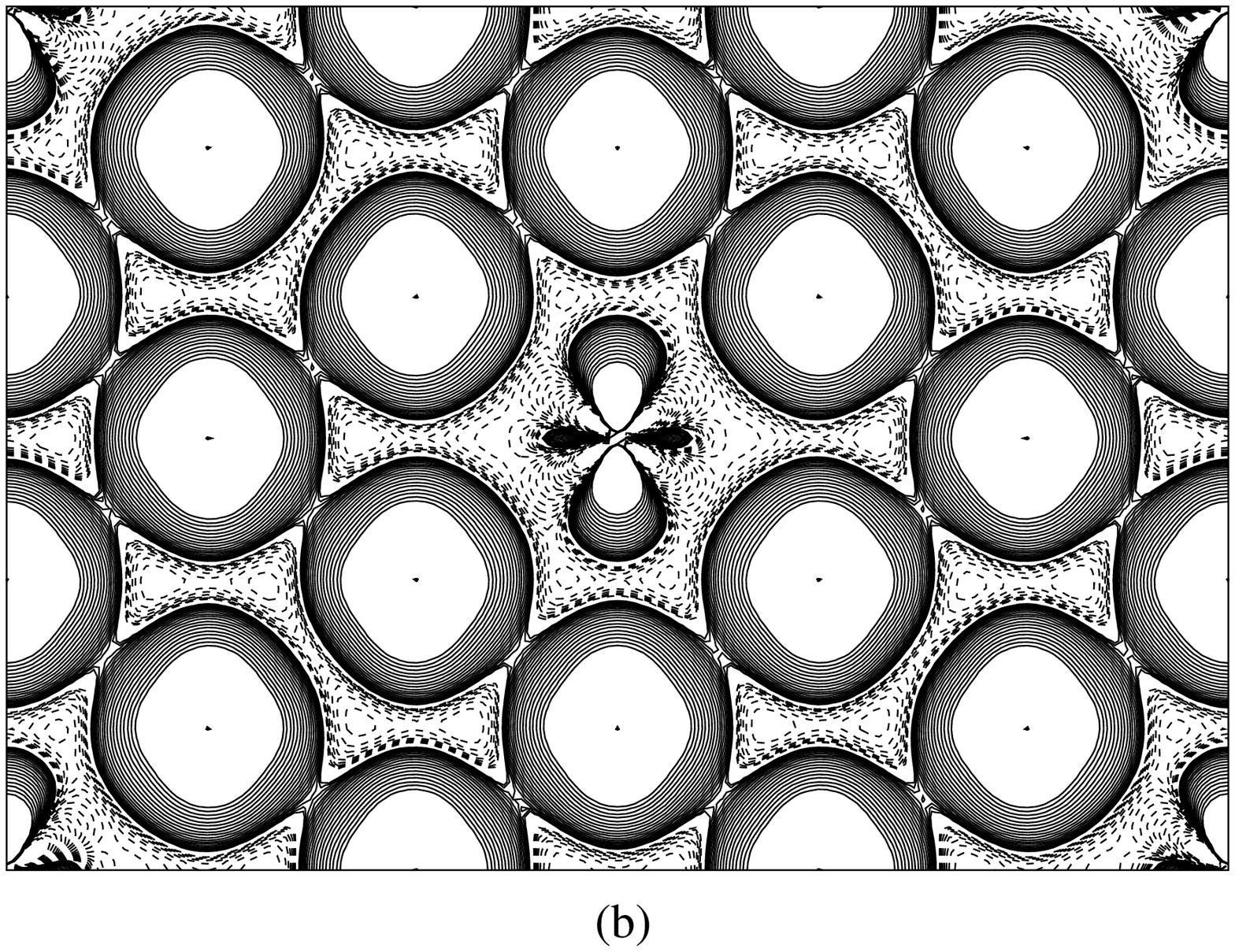}\\
\includegraphics[width=0.4\textwidth]{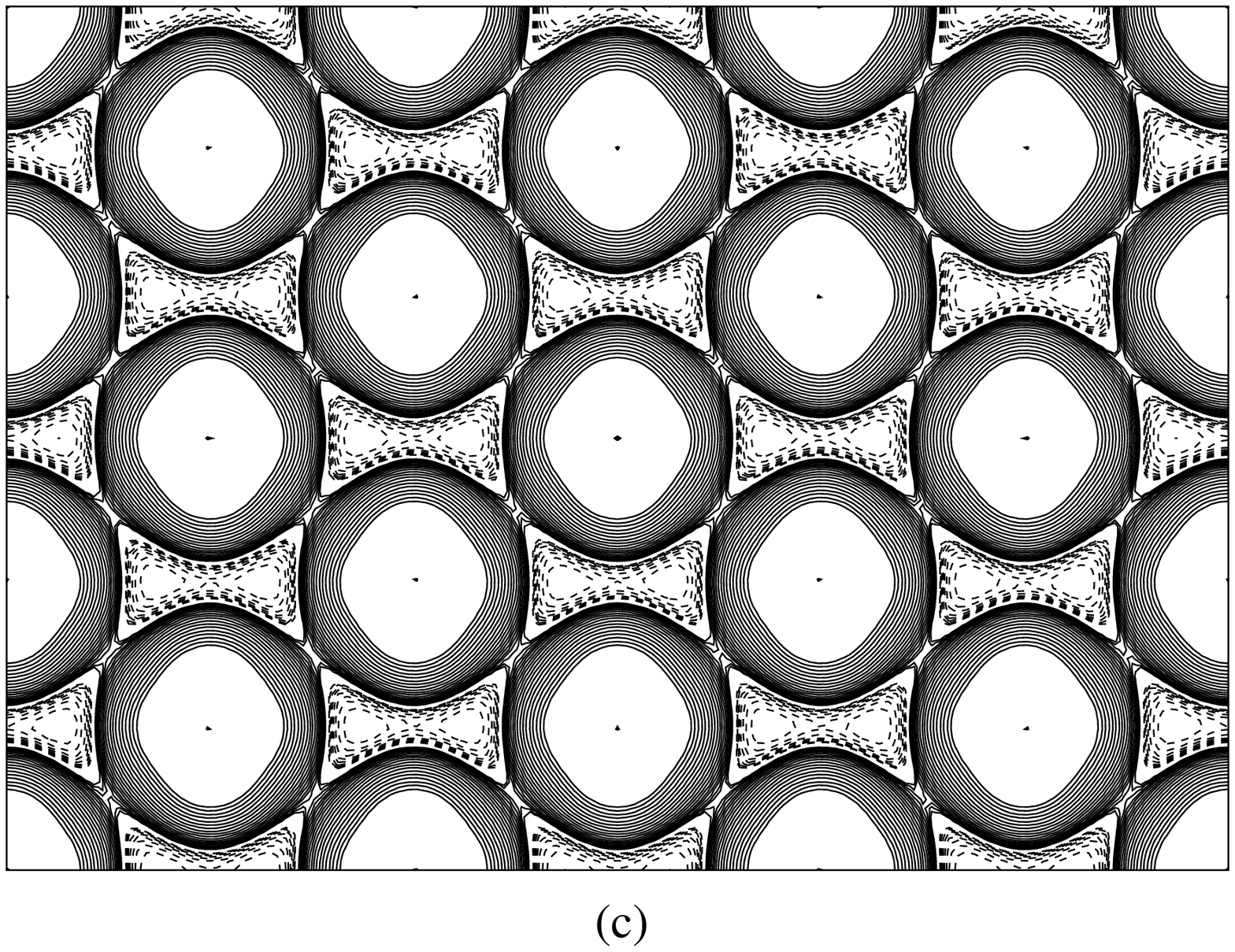}
\caption{ Calculated spin density contour plots in the (110) plane for (a) Cr, (b) Mn and 
(c) Fe inpurities in bcc Fe. 
Solid lines represent the spin-up whereas dotted lines represent 
the spin-down densities. 
The lowest contour starts from 2$\times$10$^{-4}$ electrons/a.u.$^{3}$ and
the subsequent lines differ by a factor $\sqrt{2}$. 
The Fe1, Fe2, and Fe3 atoms are also shown.} \label{spin-cont}
\end{figure}

The above mentioned facts can also be confirmed by using the calculated spin density contours. 
Representative cases (Cr, Mn, and Fe impurities in bcc Fe) are shown in Fig.~\ref{spin-cont}. 
The interstitial regions are negatively polarized. The spin density contours of pure bcc Fe show 
considerable magnetic interaction with its first neighbor, Fe1, and one can also examine the interactions 
with Fe2 and Fe3 atoms. 
This magnetic interaction is not a surprise, because the distance between the Fe1 and Fe2 atoms and 
Fe2 and Fe3 atoms is the same, \textit{i.e.}, $\sqrt{3}a/2$. 
Very recently, such kinds of interactions were also observed even for monatomic bcc Fe, 
when examined in the $(110)$ plane.\cite{topology}

The spin density at Cr site is negatively polarized and the sign reversal of the local impurity moments was 
also observable at the Mn--Cu impurity sites, 
which show positive spin polarization and couple ferromagnetically to the host Fe. 
When the $d$ bands of the impurity atom are filled, it not only affects the neighboring Fe atoms, 
but also its local impurity magnetic moment, 
and this is the case of the Zn impurity which has a small negative spin polarization (not shown here), 
the $sp$ diamagnetism.

We did not find any significant effects of SOC on $M$ as well as on the spin density contour plots 
in Fig.~\ref{spin-cont}. 
The calculated $M$ of pure bcc Fe is $\sim 2.22$\;{T} and upon SOC, $M$ is found to be $2.20$\;{T}.
Table~\ref{table1} shows that the effect of SOC is not very affective at the impurity sites, 
but slightly affects $m$ at the Fe1, Fe2, and Fe3 atoms. 
This feature is well understood by the concept of the orbital quenching, 
along with the fact of no significant Jahn-Teller distortion, \textit{i.e.} 
the interatomic distortions of the neighboring Fe atoms are negligible.\cite{mm-fe}

\subsection{Fermi contact hyperfine fields}
\label{subsec:hyperfine}

\begin{figure}
\includegraphics[width=0.8\textwidth]{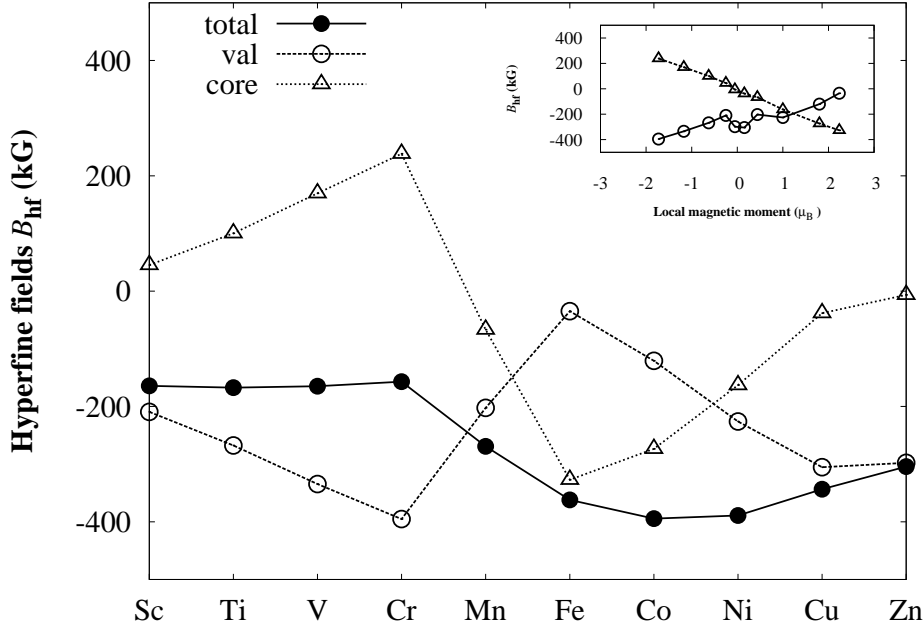}
\caption{Calculated Fermi contact hyperfine fields (in units of {kG}) of the $3d$ impurities in bcc Fe. 
The filled circles represent the total hyperfine field. 
The open circles and triangles represent the valence and core contributions, respectively. 
The inset shows the variation  of $B_{\mathrm{hf}}^{\mathrm{core}}$ 
and $B_{\mathrm{hf}}^{\mathrm{val}}$ with the local magnetic moment of the impurity atoms.} 
\label{h-fields}
\end{figure}

The hyperfine fields in bcc iron are dominated by the Fermi contact term, 
which depends essentially on the $s$-electron spin density at the nucleus; 
Fig.~\ref{h-fields}, shows both the core ($B_{\mathrm{hf}}^{\mathrm{core}}$) and valence
($B_{\mathrm{hf}}^{\mathrm{val}}$) state contributions to the fields. 
For the impurities considered, $B_{\mathrm{hf}}^{\mathrm{core}}$ and
$B_{\mathrm{hf}}^{\mathrm{val}}$ are opposite in sign with
magnitudes  increasing from Sc to Cr and then decreasing from Co to Zn. 
As a consequence there is a change of sign for the total hyperfine field in going from Mn to Fe.

The trend of $B_{\mathrm{hf}}^{\mathrm{total}}$ is similar to that for the local impurity magnetic moments. Generally,
$B_{\mathrm{hf}}^{\mathrm{core}}$ is proportional to the local impurity magnetic moment, 
as shown in the inset of Fig.~\ref{h-fields}, and its sign is negative for a parallel moment and vice versa, 
as  seen in Fig.~\ref{h-fields}. 
This linear dependence is due to the exchange interaction of 
the polarized $d$ shell with the $s$ orbitals of the core.
As a result, a weak $s$ polarization is induced at the nuclear position, 
which is typically opposite to the local magnetic moment. 
Since the exchange interaction is weak, the core hyperfine field $B_{\mathrm{hf}}^{\mathrm{core}}$ 
is expected to scale with the local moment. 
The behavior of $B_{\mathrm{hf}}^{\mathrm{val}}$ is more complicated---it scales approximately with 
the local magnetic moments but in an opposite sense to $B_{\mathrm{hf}}^{\mathrm{core}}$. 
The two major contributions to $B_{\mathrm{hf}}^{\mathrm{val}}$ are: 
(a) from the polarization of the outer $s$ orbitals of the impurity by its own local moment and 
(b) polarization of valence electrons due to the magnetic moments of the neighboring atoms 
which is usually proportional to the magnetic moment of the surrounding host atoms. 
The proportionality constants for the linear relationships described are the hyperfine coupling constants, 
calculated from Fig.~\ref{h-fields} to be $C_{\mathrm{core}} \sim -145$\,{kG$/\mu_\mathrm{B}$} and
$C_{\mathrm{val}} \sim 119$\,{kG$/\mu_\mathrm{B}$}. 
These values are comparable to the results reported for $3d$ impurities in 
Cr, Fe, and Ni hosts.\cite{Cr-hyp,Fe-hyp}

\begin{figure}
\includegraphics[width=0.8\textwidth]{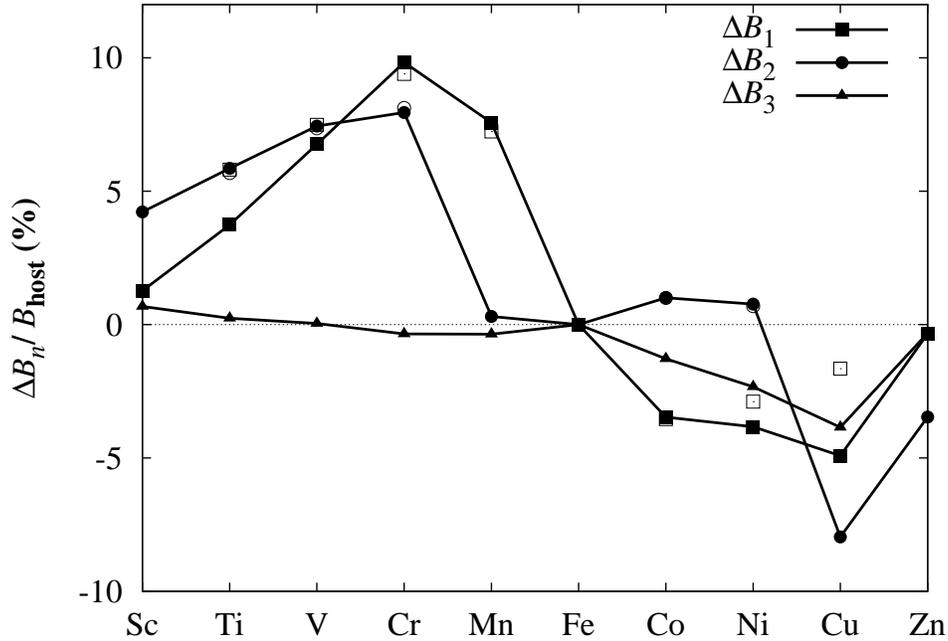}
\caption{Calculated Fermi contact hyperfine field changes $\Delta B_{n}$ normalized by 
the host Fermi contact hyperfine field $B_{\mathrm{host}}$in the 1st, 2nd and 3rd shell around 
the $3d$ impurities in bcc Fe.
Here $n$ represents the 1st, 2nd, and 3rd Fe atoms. 
The filled squares, triangles, and circles represent $\Delta B_{1}$, $\Delta B_{3}$, and 
$\Delta B_{2}$ respectively, 
whereas open symbols represent the experimental values taken from Ref.~\onlinecite{w-meinert}.} 
\label{deltah}
\end{figure}

\begingroup
\squeezetable
\begin{table}
\caption{Calculated Fermi contact hyperfine fields in units of {kG}. 
The left (right) column shows the Fermi contact field without (with) spin-orbit coupling (SOC).}
\label{table-B}
\begin{tabular}{c|ccc|cccc}
\hline\hline \multicolumn{4}{c}{Without SOC} & \multicolumn{3}{c}{With SOC} \\
\hline
Atom & Total & Valence & Core & Total & Valence & Core \\
\hline
Sc  & $-164.2$ & $-209.42$ & $\;\;\;\;54.22$ & $-163.9$ & $-209.17$ & $\;\;\;\;45.31$\\
Fe1 & $-357.3$ & $\;\;-41.95$ & $-315.34$ & $-356.3$ & $\;\;-41.51$ & $-314.82$\\
Fe2 & $-346.6$ & $\;\;-14.21$ & $-332.35$ & $-345.7$ & $\;\;-13.93$ & $-331.74$\\
Fe3 & $-359.3$ & $\;\;-36.44$ & $-322.86$ & $-358.3$ & $\;\;-36.05$ & $-322.23$\\
\hline
Ti  & $-167.3$ & $-267.54$ & $\;\;100.21$ & $-166.6$ & $-266.98$ & $\;\;104.41$\\
Fe1 & $-348.4$ & $\;\;-27.43$ & $-320.93$ & $-347.2$ & $\;\;-26.96$ & $-320.24$\\
Fe2 & $-340.6$ & $\;\;\;-8.84$ & $-331.80$ & $-339.6$ & $\;\;\;-8.60$ & $-331.01$\\
Fe3 & $-360.9$ & $\;\;-34.25$ & $-326.66$ & $-359.7$ & $\;\;-33.90$ & $-325.81$\\
\hline
V   & $-164.9$ & $-334.54$ & $\;\;169.60$ & $-164.2$ & $-333.61$ & $\;\;169.45$\\
Fe1 & $-337.4$ & $\;\;\;-8.86$ & $-328.56$ & $-336.4$ & $\;\;\;-8.23$ & $-328.14$\\
Fe2 & $-334.9$ & $\;\;\;\;-6.08$ & $-328.83$& $-334.0$ & $\;\;\;\;-5.54$ & $-328.46$\\
Fe3 & $-361.6$ & $\;\;-30.95$ & $-330.66$& $-360.5$ & $\;\;-30.35$ & $-330.17$\\
\hline
Cr  & $-157.0$ & $-395.25$ & $\;\;238.25$ & $-156.2$&$-394.07$ & $\;\;237.84$\\
Fe1 & $-326.3$ & $\;\;\;\;\;\;8.21$ & $-334.52$ & $-325.1$ & $\;\;\;\;\;\;8.98$ & $-334.11$\\
Fe2 & $-333.1$ & $\;\;\;\;-3.51$ & $-329.54$& $-332.1$ & $\;\;\;\;-3.21$ & $-328.91$\\
Fe3 & $-363.0$ & $\;\;-26.74$ & $-336.28$& $-362.0$ & $\;\;-26.42$ & $-335.56$\\
\hline
Mn  & $-269.2$ & $-202.41$ & $\;\;66.82$ & $-269.7$ & $-198.83$ & $\;\;-70.90$\\
Fe1 & $-334.5$ & $\;\;\;\;-3.19$ & $-331.29$ & $-337.7$ & $\;\;\;\;-2.73$ & $-330.95$\\
Fe2 & $-360.7$ & $\;\;-38.69$ & $-322.04$ & $-360.0$ & $\;\;-37.99$ & $-322.03$\\
Fe3 & $-363.1$ & $\;\;-32.20$ & $-330.87$ & $-362.1$ & $\;\;-31.43$ & $-330.66$\\
\hline
Fe  & $-362.0$ & $\;\;-34.84$ & $-327.12$ & $-360.2$ & $\;\;-33.48$ & $-326.74$\\
Fe1 & $-361.9$ & $\;\;-35.01$ & $-326.88$ & $-360.3$ & $\;\;-33.80$ & $-326.53$\\
Fe2 & $-361.8$ & $\;\;-34.80$ & $-327.03$ & $-360.3$ & $\;\;-33.79$ & $-326.55$\\
Fe3 & $-361.8$ & $\;\;-44.74$ & $-327.02$ & $-360.3$ & $\;\;-33.73$ & $-326.61$\\
\hline
Co  & $-394.4$ & $-120.87$ & $-273.52$ & $-493.2$ & $-120.45$ & $-272.72$\\
Fe1 & $-374.4$ & $\;\;-25.19$ & $-349.26$ & $-373.7$ & $\;\;-25.18$ & $-348.53$\\
Fe2 & $-358.2$ & $\;\;-28.75$ & $-329.44$ & $-354.4$ & $\;\;-28.58$ & $-328.86$\\
Fe3 & $-366.4$ & $\;\;-35.65$ & $-330.75$ & $-365.7$ & $\;\;-35.46$ & $-330.24$\\
\hline
Ni  & $-389.0$ & $-225.99$ & $-163.02$ & $-387.4$ & $-225.01$ & $-162.38$\\
Fe1 & $-375.7$ & $\;\;-15.49$ & $-360.25$ & $-374.8$ & $\;\;-15.29$ & $-359.48$\\
Fe2 & $-359.0$ & $\;\;-26.40$ & $-332.65$ & $-358.1$ & $\;\;-25.89$ & $-332.17$\\
Fe3 & $-370.2$ & $\;\;-33.71$ & $-336.48$ & $-369.1$ & $\;\;-33.13$ & $-335.96$\\
\hline
Cu  & $-321.3$&$-284.25$ & $\;\;-37.01$ & $-321.1$ & $-284.00$ & $\;\;-37.07$\\
Fe1 & $-366.2$&$\;\;-20.74$ & $-345.44$ & $-365.9$ & $\;\;-20.79$ & $-345.11$\\
Fe2 & $-374.6$&$\;\;-49.30$ & $-325.24$ & $-374.2$ & $\;\;-49.33$ & $-324.88$\\
Fe3 & $-361.9$&$\;\;-30.76$ & $-331.13$ & $-361.6$ & $\;\;-30.76$ & $-330.81$\\
\hline
Zn  & $-304.2$&$-297.77$ & $\;\;\;\;-6.48$ & $-303.3$ & $-296.85$ & $\;\;\;\;-6.46$\\
Fe1 & $-363.1$&$\;\;-24.89$ & $-338.23$ & $-362.1$ & $\;\;-24.56$ & $-337.55$\\
Fe2 & $-374.4$&$\;\;-46.13$ & $-328.23$ & $-373.4$ & $\;\;-45.81$ & $-327.62$\\
Fe3 & $-362.9$&$\;\;-29.95$ & $-332.96$ & $-362.0$ & $\;\;-29.67$ & $-332.34$\\
\hline\hline
\end{tabular}
\end{table}
\endgroup

Figure~\ref{deltah} illustrates the changes, $\Delta B_{1}$, $\Delta B_{2}$, and $\Delta B_{3}$ 
in the Fermi contact hyperfine fields of the Fe1, Fe2, and Fe3 atoms, respectively, 
with the normalization by the host Fermi contact hyperfine field, $B_{\textrm{host}}$. 
Positive values imply that the hyperfine field of Fe${n}$ neighbor is smaller than that of the host and vice versa. 
The calculated $\Delta B_{1}$ for Ni and Co are negative, in agreement with experiments,\cite{w-meinert} 
whereas $\Delta B_{1}$ is inconsistent with the calculations done by Dederichs \textit{et al.},\cite{w-meinert} 
who found a zero value for $\Delta B_{1}$. 
This discrepancy is due to partly the different computational method. 
Our calculated GGA trend of $\Delta B_{n}$ can be compared with the experimental observations and 
the other previous  calculations.\cite{w-meinert}

The $B_{\mathrm{hf}}$ results in Table~\ref{table-B} show that
$B_{\mathrm{hf}}^{\mathrm{val}}$ of Fe1 is negative for all the $3d$ impurities 
except for Cr, and these positive quantities increase the transfer field of the Fe1 atoms 
through its proportionality to the  magnetic moment of the surrounding atoms, 
\textit{i.e.}, the Fe1, Fe2, and Fe3 atoms. 
Hence, the transfer field of the Fe2 atom is smaller than that of the Fe1 atom and 
this supports  the notion that the reduced $M$ of bcc Fe alloyed with Cr and Mn is caused mainly by 
the Fe2 atoms.

Reported $B_{\mathrm{hf}}$ values for pure bcc Fe calculated using LSDA and GGA are, 
without SOC, $-278$\;{kG} and $-298$\,{kG}, respectively, and, 
with SOC, $-253$\,{kG} and $-276$\,{kG}, respectively.\cite{LDA-hyp} 
Our GGA result is $-371$\,{kG} (and $-364$\,{kG} with SOC), 
which is close to the experimental value of $-339.0$\,{kG}.\cite{Fe-hyp-expt} 
The somewhat larger values obtained here are consistent with FLAPW-LSDA value of 
$-366.0$\,{kG} reported by Ohnishi.\cite{Ohnishi} 
Table~\ref{table-B} shows again that SOC does not significantly affect the calculations 
because of the localization of the impurity.

Finally, some test calculations were also carried out for big and small supercells of Fe$_{52}$Ni$_{2}$ and
Fe$_{15}$Ni and the local properties around Ni were similar to Fe$_{26}$Ni. 
This indicateds that $3\times 3\times 3$ supercell of the primitive bcc cell as discussed in 
Sec.~\ref{sec:method} is sufficient for $3.7$\,{at.\%} impurities in bcc Fe, 
but may not be sufficient for the other host, \textit{e.g.} bcc Cr.\cite{Cr-hyp}

\section{Summary}
\label{sec:summary}

The magnetism and electronic structures of $3d$ impurities in body-centered iron have been investigated 
in terms of the FLAPW method based on the generalized gradient approximation. 
The results compare favorably with existing experimental data, 
and have been contrasted with published calculations where available.

The lattice parameters were optimized and it is shown that the lattice constant decreases with 
the atomic number of the impurity $X$ (Sc--Fe), 
whereas it increases for Co--Zn. 
The calculated formation energy results are good agreement with the alloying experiences in steels design. 
It is found that iron induces a magnetic moment at the Sc, Ti, V, Cu, and Zn atoms substituted into the iron lattice; 
Sc, Ti, V, Cr, and  Zn  couple antiferromagnetically with iron, 
resulting in an overall reduction in the saturation magnetization relative to pure iron. 
In contrast, Co, and Ni couple ferromagnetically with iron and resulting in increase in saturation magnetization. 
The trend of the atom averaged magnetic moment versus effective atomic number is shown 
to lie on the left branch of the well-known Slater-Pauling curve, but it is not simply linear---the deviation from 
the standard Slater-Pauling curve is attributed by the nature of the weak ferromagnetism of bcc iron. 
These results have been interpreted by examining in detail the local density of states, near neighbor interactions, 
and other features of the electronic structure of the alloys resulting from the substitution of impurity atoms into iron. 

An important outcome is that calculations of the type reported here are not significantly influenced by
spin-orbit coupling. 
The calculated Fermi contact hyperfine fields also support this conclusion. 
In the light of previous theoretical calculations, where LSDA was used our calculations showed that 
although LSDA can not describe the correct magnetic properties of bcc Fe, 
but can describe the correct magnetic properties (qualitatively) of bcc Fe because our GGA results are close to 
the previous LSDA results. 
Comparing the LSDA and GGA, we showed that the GGA results are closer to the experimental data.

\begin{acknowledgments}
This work was supported by the Steel Innovation Program by POSCO, the WCU (World Class University) program
(Project No. R32-2008-000-10147-0), the Basic Science Research Program (Grant No. 2009-0088216) through 
the National Research Foundation funded by
the Ministry of Education, Science and Technology of Republic of Korea. 
The present work was also funded by office of Scientific Research (Grant No. FA9550-07-1-0013) and 
in part by the U.S. Air Force.

We are grateful to Professor Hae-Geon Lee for the provision of laboratory facilities at POSTECH. 
Finally, IGK thanks Prof. Sam Kyu Chang for encouraging the initiation of this work.
\end{acknowledgments}

\end{document}